\documentclass[letterpaper, 10 pt, conference]{ieeeconf}  

\usepackage{graphics}
\usepackage{graphicx, tabularx}
\usepackage{subfigure}
\usepackage[colorinlistoftodos]{todonotes}
\usepackage[colorlinks=true, allcolors=blue]{hyperref}

\IEEEoverridecommandlockouts                              

\title{\LARGE \bf
On early brain folding patterns using biomechanical growth modeling
}

\author{Xiaoyu Wang$^{1}$, Amine Bohi$^{2}$, Mariam Al Harrach$^{3}$, Mickael Dinomais$^{3}$, Julien Lef\`evre$^{2}$, Fran\c{c}ois Rousseau$^{1}$
\thanks{$^{1}$IMT Atlantique, LaTIM U1101 INSERM, UBL, Brest, France.~{\tt\small xiaoyu.wang@imt-atlantique.fr, francois.rousseau@imt-atlantique.fr}}%
\thanks{$^{2}$Aix Marseille Univ, CNRS, INT,~Inst Neurosci~Timone,~Marseille,~France. 
    }%
\thanks{$^{3}$LUNAM, Universit\'{e} d'Angers, LARIS - EA7315, F-49000, Angers, France.
    }%
}

\begin{document}

\maketitle
\thispagestyle{empty}
\pagestyle{empty}

\begin{abstract}

Abnormal cortical folding patterns may be related to neurodevelopmental disorders such as lissencephaly and polymicrogyria. In this context, computational modeling is a powerful tool to provide a better understanding of the early brain folding process. Recent studies based on biomechanical modeling have shown that mechanical forces play a crucial role in the formation of cortical convolutions. However, the correlation between simulation results and biological facts, and the effect of physical parameters in these models remain unclear. In this paper, we propose a new brain longitudinal length growth model to improve brain model growth. In addition, we investigate the effect of the initial cortical thickness on folding patterns, quantifying the folds by the surface-based three-dimensional gyrification index and a spectral analysis of gyrification. The results tend to show that the use of such biomechanical models could highlight the links between neurodevelopmental diseases and physical parameters.

\end{abstract}

\section{INTRODUCTION}
\label{sec:intro}
Brain growth is accompanied by the folding of the cerebral cortex, which takes place during gestational weeks 16-40~\cite{sun2014growth}. Recent studies have revealed that not only the molecular and cellular processes but also mechanical forces play an important role in the formation of the gyral and sulcal convolutions. For the latter, computational modeling is a key point to further understand brain folding mechanism. Based on a combination of two hypotheses for cortical folding, which are axonal tension and differential growth, very preterm neonatal brain growth is modeled by a morphogenetic growth of the cortical layer and a stretch-driven growth of the subcortical layer \cite{budday2014mechanical}. The model and 2D simulation results show that mechanical stretch plays a crucial role in human brain development. Another human cerebral folding model (e.g., \cite{tallinen2016growth,tallinen2014gyrification}), which is based on the hypothesis of differential tangential growth, can produce the characteristic patterns of convolutions on the brain surface over a developmental process. This model is guided by the use of 3D magnetic resonance images of a smooth fetal brain as an initial point. The 3D numerical simulations of the brain demonstrate that the relative tangential expansion of the cerebral cortex constrained by the white matter generates compressive stress, resulting in cusped sulci and smooth gyri similar to those in developing fetal brains. 

However, the correspondence concerning the increase in brain volume, the development of the folding patterns and gestational age is not fully realistic in these models. To make the increase in the brain volume more in agreement with the biological measurement results (e.g., \cite{chang2003assessment,hsu2013quantitative,armstrong1995ontogeny,gong1998fetal}), we propose a new brain longitudinal length (BLL) growth model, which is integrated into the human cortical folding model \cite{tallinen2016growth}, to improve the brain volume accuracy during the cerebral folding process. 

The cortical folding patterns are influenced by various physical parameters, e.g., the initial cortical thickness \cite{budday2014mechanical}, the initial geometry \cite{verner2017computational} and the initial curvature of the surface \cite{kroenke2018forces}. A deeper understanding of brain physical parameters can significantly contribute to comprehend certain pathologies associated with abnormal cortical folding processes, which motivates us to study the effect of the initial cortical thickness on cortical folding patterns. In this study, we build numerical simulations of different initial cortical thicknesses using our combined model, which first allow us to visually remark the difference of the folding patterns. Then, we quantify these folding patterns by the surface-based three-dimensional gyrification index and the curvature-based spectral analysis \cite{germanaud2012larger} to compare the complexity of the folding patterns.

\section{METHODS}
\label{sec:format}

\subsection{Biomechanical model of brain folding}
\label{ssec:biomeca}
In (e.g., \cite{tallinen2016growth,tallinen2014gyrification}), the authors propose a human cortical folding model, which can mimic a realistic brain folding process. The brain is modeled as a neo-Hookean hyperelastic material solid. Brain growth is modeled by a relative tangential expansion of the cortical layer and the white matter layer, the cortical layer is assumed to grow more rapidly than the white matter layer. The brain model is discretized into high-density tetrahedral finite elements for the simulations of the developing brain.

The initial condition for the growth of this brain model is derived from the 3D MRI of a 22 weeks' fetal brain, with the displacement and the velocity which are zero. The model obeys the Neumann boundary condition, in other words, the force is imposed on the brain model in finite element calculation. There are two main forces considered in the model. One is the elastic force, which is derived from the volumetric strain energy density of neo-Hookean and a deformation gradient. Another is the contact force, which takes place when a separation between a node and a triangle face at the brain surface is less than a threshold in order to prevent nodes from penetrating element faces.

The deformation gradient is defined in this model by $F = A(G\hat{A})^{-1}$, which differs from the traditional definition of $F = A\hat{A}^{-1}$, where $A$ is the deformed configuration of a tetrahedron, $\hat{A}$ is the stress-free initial configuration of a tetrahedron. $G$ is the relative tangential growth tensor, which associates with the distance of a tetrahedron from surface in material coordinates. Thus, when the relative tangential growth tensor is initialized, the brain model starts to grow and the deformation gradient is formed, the corresponding elastic force can be calculated. The contact force at the brain surface is obtained via penalty based vertex-triangle contact processing \cite{ericson2004real}. The resultant force is applied as the nodal force on each node of the mesh to produce the deformation of the brain model.

\textbf{BLL growth model} In order to explain the brain volume growth during the cortical folding process, Tallinen \textit{et al.} propose a brain longitudinal length ($BLL$) growth model which is defined by $BLL~(mm) = 59/(1-0.55t)$, where $t$ parametrizes time and has a non-linear relation to gestational age ($GA$) as $t = 6.926\times10^{-5}GA^3-0.00665GA^2+0.250GA-3.0189$. $t\in[0,1]$ corresponds to $GA\in[22 weeks, adult]$. According to the longitudinal length, the size of the initial brain will be scaled isotropically (in three dimensions). After measuring the brain volume of the simulation results, we find that this BLL growth model lacks brain volume growth during the folding process and cannot mimic a volume growth process consistent with the biological measurement results (e.g., \cite{chang2003assessment,hsu2013quantitative,armstrong1995ontogeny,gong1998fetal}). Here, we propose a new BLL growth model. Referring to the normative fetal brain longitudinal length data in \cite{kyriakopoulou2017normative}, we find that a second-order polynomial is good enough to fit them and thus define the brain longitudinal length (BLL) as a function of gestational age (GA) from 22 to 40
weeks as
\begin{equation}
BLL~(mm) = -0.067GA^2 + 7.16GA - 66.05, 
\end{equation}
with $R^2 = 0.9987$. The brain mesh is then scaled isotropically according to the longitudinal length of each gestational age. 


In the cortical folding model (e.g., \cite{tallinen2016growth}, \cite{tallinen2014gyrification}), the relative tangential growth tensor is calculated for each tetrahedron by $G = gI+(1-g)\hat{n}\otimes{\hat{n}}$,
which is perpendicular to the normal vector $\hat{n}$ of the tetrahedron. $g$ is the expansion ratio of the gray matter relative to the white matter, which is defined by $g = 1 + \frac{\alpha}{1+\exp(10(\frac{y}{H}-1))}$, where $\alpha$ controls the magnitude of expansion and is defined as $\alpha = (\sqrt{8}-1)t$~\cite{tallinen2016growth}. $y$ is the distance from the top surface, which is calculated for four vertices of each tetrahedron and will be averaged. $H$ is the cortical thickness. Considering that the cortical thickness changes little and increases steadily during the expanding process, we thus also propose to model the cortical thickness~$H$ as a linear function of time, as follows: $H = H_i + \beta{t}$, where $H_i$ is the initial cortical thickness and $\beta$ determines the magnitude of thickening. 
For a typical cortical thickness of 22 weeks' fetal brains, which is 2.5 mm, the corresponding deformed cortical thickness is approximately 2.8 mm in adulthood ($t$ = 1), we thus define $\beta$ = 0.3 in our simulations.

Then we combine the cortical folding model, the new BLL growth model and the model of the cortical thickness variation. Based on this 
combined model, we study in this work the effect of the initial cortical thickness on folding patterns. 

\subsection{Spectral analysis of gyrification}
\label{ssec:Spangy}
To quantify the cortical folding patterns of our simulations, we adopt the method for the spectral analysis of gyrification \cite{germanaud2012larger}. This method performs a spectral decomposition of the mean curvature of the brain surface mesh based on the Laplace-Beltrami operator eigenfunctions. It allows to extend Fourier Analysis to more general domains such as graphs or surfaces and to produce power spectra. The Laplace-Beltrami operator is defined as $\Delta_{M} = div\cdot\nabla_{M}$, where $M$ is a Riemannian manifold. The eigenvalues of $-\Delta_{M}$ are $\lambda_{0} = 0\leq\lambda_{1}\leq\ldots$ and $\phi_{0}$, $\phi_{1}$, $\ldots$ are associated orthonormal basis of eigenfunctions, which satisfy
\begin{equation}
-\Delta_{M}\phi_{i} = \lambda_{i}\phi_{i}.
\end{equation}
The approach of the decomposition of the mean curvature ($C$) is described by
\begin{equation}
C = \sum\limits_{i=0}^{+\infty}C_{i}\phi_{i},  C_{i} = \int_{M}C\phi_{i},
\end{equation}
where $C_{i}$ is the Fourier coefficient of the curvature in the eigenfunctions basis $\phi_{i}$.
A band power spectrum is defined by
\begin{equation}
BS_{C}(0) = C^{2}_{0}, BS_{C}(k) = \sum\limits_{i=i^{k}_{1}}^{i^{k}_{2}}C^{2}_{i},
\end{equation}
where $k$ is the number of the band, B0 power corresponds to the first constant eigenfunction, other bands correspond to sums of powers in different frequency bands that defined via the wavelengths. $i^{k}_{1}$ and $i^{k}_{2}$ correspond to the numbers of eigenfunctions associated to the band interval limits.

The approach of the local dominant band segmentation, which is presented in \cite{germanaud2012larger}, allows us to parcellate the cortical surface into 7 spatial frequency bands, from B0 to B6. The bands B1, B2 and B3 are related to global brain shape. The last three bands (B4, B5 and B6) are associated with fold-related variations of curvature, which reveal the number of 1st, 2nd and 3rd order folds, respectively. The 3rd order folds locate where the absolute values of curvature are the biggest on the brain surfaces. In consideration that we want to quantify the folding variations induced by the difference in the initial cortical thickness, we focus thus on the folding power of B4, B5 and B6.

\subsection{Surface-based three-dimensional gyrification index}
\label{ssec:GI}
In order to represent quantitatively the degree of cortical convolution, we use the surface-based three-dimensional gyrification index (3D GI), which is a global measure and defined as the ratio of the cortical surface area to the area of its smooth 'convex hull' (the minimum surface area needed to completely enclose the brain) \cite{clouchoux2012quantitative}. For the simulations of different initial cortical thicknesses, we calculate and compare the 3D GI, which will be shown in the results section.

\section{RESULTS}
\label{sec:results}

\subsection{Validation of brain volume with biological time}
\label{ssec:valvolume}
\begin{figure}[ht!]
\centering
\includegraphics[scale=0.48]{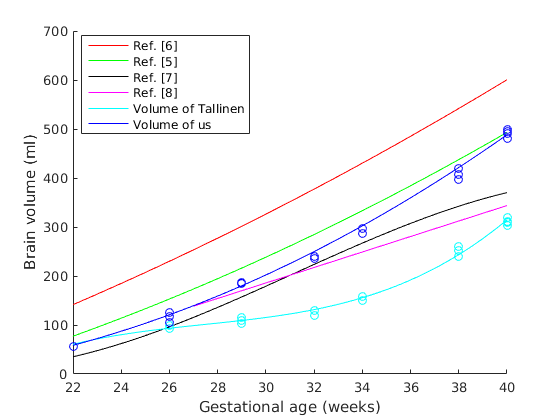}
\caption{\label{fig:volume}The comparison of the brain volume development between our simulation results, simulations results of Tallinen's model and literature data.}
\end{figure}

Several numerical simulations of a 22 weeks fetal brain have been performed by using the combined biomechanical model described in Section~\ref{ssec:biomeca}. We have then computed the fetal brain volume of the simulations with the initial cortical thicknesses varying from 1.48 to 2.98 mm. A comparison between the volume data of our BLL growth model, the volume data using Tallinen's BLL growth model and other validated data of the literatures (e.g., \cite{chang2003assessment,hsu2013quantitative,armstrong1995ontogeny,gong1998fetal}) is shown in Fig.\ref{fig:volume}. Comparing with the volume data of Tallinen (cyan points), our volume data (blue points), fitted with a second-order polynomial: $BV~(ml) = 0.59GA^{2} - 12.77GA + 53.46$, is much more consistent with other validated brain volume data of the literatures. At the same gestational age, the different points represent the brain volume of the simulations with different initial cortical thicknesses.

The blue fitting curve of our volume data is consistent with these validated brain volume curves during gestational weeks, showing that our brain longitudinal length (BLL) growth model can mimic a realistic brain volume growth process.

\subsection{Impact of cortical thickness on surface morphology}
\label{ssec:initcortical}

\begin{figure}[ht!]
\centering
\includegraphics[scale=0.41]{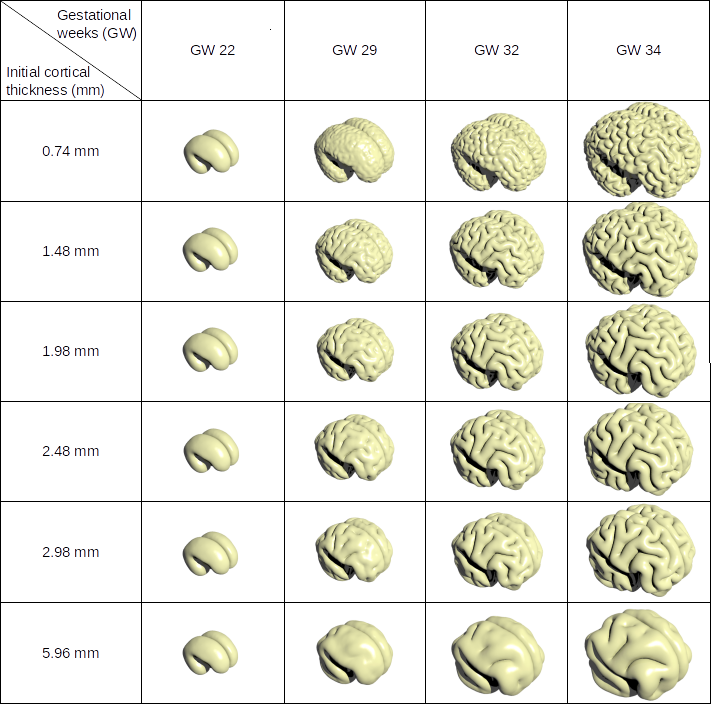}
\caption{\label{fig:cormorpho}Cortical morphology with increasing gestational weeks for different initial cortical thicknesses.}
\end{figure}

To understand the effect of the initial cortical thickness on the cortical folding patterns, we vary the initial cortical thickness in the combined model from 0.74 to 5.96 mm. The cortical thicknesses from 1.48 to 2.98 mm are defined according to normative fetal brain measurements. The other two cortical thicknesses (0.74 and 5.96 mm) are our hypotheses for fetal abnormal cortical thicknesses. 
The simulation results are shown in Fig.\ref{fig:cormorpho}. At the 29th week of gestation, the brain surfaces already showed some clear line-like sulci. As the cortical thickness increases, the number of sulci decreases, the sulci became more isolated. Starting from 32 GW, smooth gyri begin to appear on cortical surface. With the increase in the cortical thickness, the gyri became wider and fewer. The qualitative analysis is consistent with previous works such as \cite{budday2014mechanical} relating the cortical thickness and the folding patterns. A greatly increased cortical thickness associates with a severe decrease in number of gyri and sulci, which corresponds to the phenomenon of lissencephaly. The morphology with the initial cortical thickness of 5.96 mm in our simulations is similar to lissencephaly. On the contrary, polymicrogyria relates to an overly convoluted cortex, i.e., an increase in number and a decrease in size of gyri and sulci. The morphology with the thin initial cortical thickness of 0.74 mm resembles polymicrogyria.

\begin{figure}[ht!]
\centering
\includegraphics[scale=0.48]{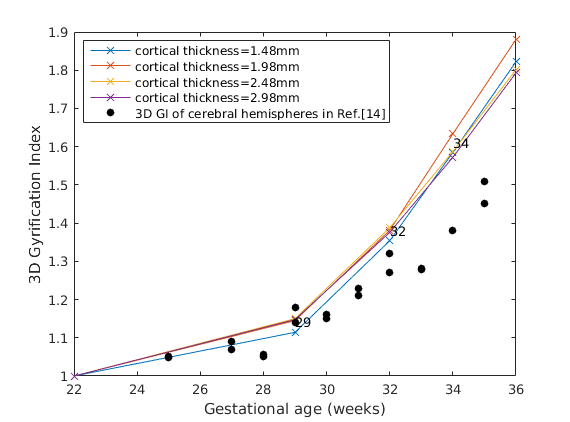}
\caption{\label{fig:3DGI}3D gyrification index computed on each brain surface.}
\end{figure}

To analyze quantitatively the folding patterns, we first compute the 3D GI on the surfaces of the simulated brain (the initial cortical thicknesses are set between 1.48 and 2.98 mm), as is shown in Fig.\ref{fig:3DGI}. It can be seen that there is a clear increase of the 3D GI with the increase in gestational age, regardless of the cortical thickness. The difference in the 3D GI is small at the same gestational week. However, the effect of the cortical thickness on the 3D GI is irregular. The GI based on surface areas may not be able to describe the complexity of the sulci and gyri when the initial cortical thickness changes little. The 3D GI computed on cerebral hemispheres for 12 healthy fetuses \cite{clouchoux2012quantitative} is plotted for comparison. We remark a good agreement on the 3D GI at early gestational ages, but bias appears at later gestational ages may be due to differences in individual brain development. In future work, we will use more fetal brain MRI data of different gestational ages and compare the simulation results with the data to further validate our model.

Secondly, for different initial cortical thicknesses ($H_i$ varying from 0.74 to 5.96 mm), we analyze the curvature spectra and compute the folding power of B4, B5 and B6 on the brain surfaces. The total folding power is shown in Table \ref{tab:totpower}. At the same gestational week, increasing the initial cortical thickness decreases the amount of growth in cortex which results in smaller total folding power, i.e., less total folds (1st, 2nd and 3rd order folds).

\begin{table}[ht!]
\caption{\label{tab:totpower}Total folding power of B4, B5 and B6}
\centering
\begin{tabular}{ | c | c | c | c | c | }
\hline
    & GW 22 & GW 29 & GW 32 & GW 34 \\
    \hline
    $H_i$ = 0.74 mm & 3.76 & 73.11 & 325.59 & 854.95 \\
    $H_i$ = 1.48 mm & 3.76 & 59.26 & 177.48 & 366.06 \\
    $H_i$ = 1.98 mm & 3.76 & 50.89 & 125.71 & 272.16 \\
    $H_i$ = 2.48 mm & 3.76 & 39.06 & 94.39 & 194.27 \\
    $H_i$ = 2.98 mm & 3.76 & 33.39 & 73.80 & 152.26 \\
    $H_i$ = 5.96 mm & 3.76 & 15.48 & 19.49 & 29.01 \\
    \hline
\end{tabular}
\end{table}

\begin{figure}[ht!]
\centering
\includegraphics[scale=0.48]{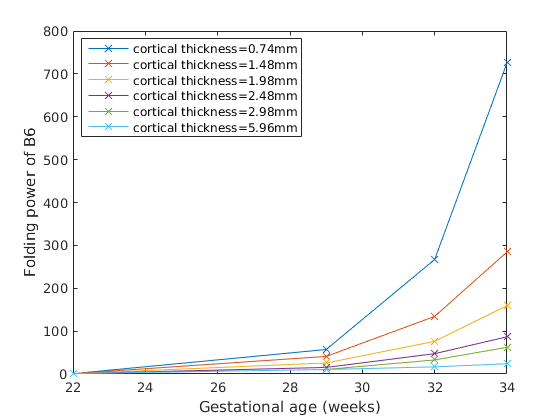}
\caption{\label{fig:power}Folding power of B6 for different cortical thicknesses.}
\end{figure}

Furthermore, we find that, for the thinner cortical thicknesses, the folding power of B6 accounts for an increasingly large proportion of the total folding power with the increasing gestational age. At each gestational week, the difference in B6 folding power is larger than that in B4 and B5 folding power between these cortical thicknesses, indicating that the small changes in cortical thickness have significant effect on the number of tertiary brain folds. These trends of B6 folding power for different cortical thicknesses are shown in Fig.\ref{fig:power}. The thinner the initial cortical thickness, the larger the increase of the B6 folding power demonstrates. 

\section{CONCLUSION}
\label{sec:discuss}
In this work, we have proposed to introduce a brain longitudinal length growth model derived from normative fetal brain measurement into a biomechanical cortical folding model. This has led to more realistic early brain development simulations validated by previous works of the literature. Furthermore, we have evaluated the impact of the initial cortical thickness onto the folding patterns through visual analysis and quantitative features. More specifically, experiments have shown that thinner cortical thickness leads to higher B6 folding power that corresponds to tertiary brain folds. 
Such biomechanical model may be used for future works 
to investigate causes of neurodevelopmental diseases associated with abnormal folding patterns of the cerebral cortex.

\addtolength{\textheight}{-12cm}   




\section*{ACKNOWLEDGMENT}

The research leading to these results has been supported by the Fondation pour la Recherche M\'edicale (grant DIC20161236453).


\bibliographystyle{ieeetr}
\bibliography{IEEEexample}

\end{document}